# Collective Behavior of Crowded Drops in Microfluidic Systems

Ya Gai[1], Andrea Montessori[2], Sauro Succi[3,4], and Sindy K.Y. Tang[5*]

[1] 319 Stanworth Ln, Princeton, NJ 08540, USA

[2] Dipartimento di Ingegneria, Università degli Studi Roma TRE, via Vito Volterra 62, Rome, 00146, Italy

[3] Center for Life Nano&Neurosciences @ La Sapienza, Italian Institute of Technology, via Regina Elena, 259, 00161, Roma, Italy, and

[4] Physics Department, Harvard University, Oxford Street 29, MA 02138, Cambridge, USA

[5] Department of Mechanical Engineering, Stanford University, Stanford, CA 94305, USA

*sindy@stanford.du

**Abstract**

Droplet microfluidics, in which micro-droplets serve as individual reactors, has enabled a wide range of high-throughput biochemical processes. Unlike solid wells typically used in current biochemical assays, droplets are subject to instability and can undergo breakup, especially under fast flow conditions. Although the physics of single drops has been studied extensively, the flow of crowded drops or concentrated emulsions—where droplet volume fraction exceeds ~80%—is relatively unexplored in microfluidics. In this article and the related invited lecture from the 74[th] Annual Meeting of the American Physical Society's Division of Fluid Dynamics, we describe the collective behavior of drops in a concentrated emulsion by tracking the dynamics and the fate of individual drops within the emulsion. At the slow flow limit of the concentrated emulsion, we observe an unexpected order, where the velocity of individual drops in the emulsion exhibits spatiotemporal periodicity. Such periodicity is surprising from both fluid and solid mechanics points of view. We show the phenomenon can be explained by treating the emulsion as a soft crystal undergoing plasticity, in a nanoscale system comprising thousands of atoms as modeled by droplets. Our results represent a new type of collective order not described before and have practical use in on-chip droplet manipulation. As the flow rate increases, the emulsion transitions from a solid-like to a liquid-like material, and the spatiotemporal order in the flow is lost. At the fast flow limit, droplet breakup starts to occur. We show that droplet breakup within the emulsion follows a probability distribution, in stark contrast to the deterministic behavior in classical single-drop studies. In addition to capillary number and viscosity ratio, break-up probability is governed by a confinement factor that measures drop size relative to a characteristic channel length. The breakup probability arises from the time-varying packing configuration of the drops. Replacing surfactant with nanoparticles as droplet stabilizers suppresses break-up and increases the throughput of droplet processing by >300%. Strategic placement of an obstacle suppresses break-up by >$10^3$ fold. Finally, we discuss recent progress in computation methods for recapitulating the flow of concentrated emulsions.

# 1. Introduction

Understanding the flow of droplets in highly confined, microscale devices is key to the design and application of droplet microfluidics [1–7]. Droplet microfluidics uses monodisperse droplets, with volumes from picoliter to nanoliter, as individual reactors for biochemical reactions. The droplets are typically aqueous suspended in an immiscible oil and can contain chemical and/or biological species, including nucleic acids, proteins, cells, and even whole organisms. To prevent coalescence, the droplets are stabilized by a surfactant [8–19]. Droplet technology has enabled massive parallelization of reactions with sub-nanoliter reagent consumption per reaction [1,2,4–6,20–27] and has been shown to increase the throughput of certain assays up to $10^3$x from industry standards using multi-well plates [25,28–31]. Droplet microfluidics has already found commercial success in the biotechnology industry for processes such as digital polymerase chain reaction and single-cell sequencing [32,33].

In various processing steps in droplet microfluidics, such as the incubation and interrogation steps, the droplets can become crowded with a volume fraction of the disperse phase (i.e., drops) exceeding 80%, forming a high internal phase emulsion or concentrated emulsion. Concentrated emulsions have vast industrial applications beyond droplet microfluidics, including polymer manufacturing, oil recovery, food processing, and cosmetic production [34–41]. The rich mechanical responses of emulsions and other complex fluids, including colloids, granular media, and foams, are central to soft matter physics and materials science [42–46].

The physics of droplets and emulsions has a long history. The dynamics of a single droplet has been extensively studied since the pioneering work by G.I. Taylor in the 1930s [47,48]. The detailed descriptions, both from theoretical and experimental perspectives, can be found in classical papers and reviews [49–56]. The study of single droplets cannot be applied directly to concentrated emulsions, however, due to the missing interactions among droplets. Above the random close packing limit ($\varphi_{RCP}$ ~64% for monodisperse spherical particles), the droplets in the emulsion form a fragile network that shows a rich and complex mechanical response. The response can be either solid-like or liquid-like, depending on the externally applied stress [52–54]: below the yield stress, the emulsion does not flow and exhibits an elastic response that is almost linear; above the yield stress, the emulsion starts to flow and exhibits a viscous, strain-rate-dependent response. Previous work on concentrated emulsions has primarily focused on their bulk rheology [57–64]. However, the fate of individual droplets within a flowing concentrated emulsion remains unexplored. Understanding how individual droplets behave within a concentrated emulsion is critical to droplet microfluidics because each droplet can contain a different biochemical species and corresponding reaction.

Motivated by this knowledge gap, our work aims to investigate how the emulsion flows at the individual droplet level within a microfluidic system (Fig. 1). The overall question we ask is: *in a crowded and confined system where many drops interact, how do the interactions give rise to collective behavior?* Given the application of these emulsions in droplet microfluidics, we are also interested in the following engineering questions: *How do many-drop interactions impact the performance of droplet technology? How can we overcome or leverage such interactions to enhance the performance of droplet technology?*

Our experimental work focuses on a two-dimensional (2D) emulsion flowing as a monolayer because they are the most relevant to droplet microfluidic applications. They are also simple to monitor with standard microscopy and a high-speed camera. Most of the microchannels we have studied have a slowly contracting or tapered geometry, where the width of the channel decreases from ~5-10 droplet diameters to <1 droplet diameter at the constriction. This tapered geometry is often used for the serial interrogation of droplets, sometimes followed by droplet sorting based on the encapsulated content. The tapered geometry forces the droplets to rearrange, failing which, the droplets might pinch off each other to sustain the flow.

The following sections are organized as follows. Sections 2-5 are organized based on the strain rate of emulsion flows characterized by the capillary number ($Ca$). Sections 2-3 fall in a low capillary number regime ($Ca \sim 10^{-4}$), where interfacial effects dominate the viscous effects. We examine an unexpected order in the rearrangement of the drops that could be explained by treating the drops as a microfluidic crystal. Section 4 discusses the transition from a low $Ca$ to a moderate $Ca$ regime. Sections 5-6 cover a high capillary number regime ($Ca > 10^{-2}$) where viscous effects become important, and emulsions start to flow laminarly and can exhibit unstable phenomena such as droplet breakup. Section 7 summarizes recent efforts in computational modeling of concentrated emulsions.

## 2. Micro-PIV study on the internal flow of concentrated emulsion droplets in microchannels in the low *Ca* regime

This section examines the flow inside individual droplets within a concentrated emulsion flowing as a monolayer in microchannels. Understanding flow patterns inside the drops is important for predicting the degree of mixing critical for biochemical reactions.

Fig. 2a shows the µPIV setup [65,66]. Images were acquired at the mid-height of the microchannel with an inverted microscope and a high-speed camera. We started with a simple microchannel geometry – a narrow, straight microchannel in which droplets spanned the width of the channel in multiple parallel rows (Fig. 2b). When the emulsion flowed as a single row of droplets from left to right in the channel, we observed two co-rotating structures in each half of the droplet (see [66] for details). The top half had clockwise rotating structures, while the bottom half had counterclockwise rotating structures. The presence of co-rotating structures is further confirmed by calculating the *Q*-criterion as $Q = \frac{1}{2}(\|\Omega\|^2 - \|S\|^2)$ [67], where $S$ and $\Omega$ are the symmetric and antisymmetric parts of the velocity gradient tensor $\nabla u$, respectively (i.e., $S = \frac{1}{2}(\nabla u + (\nabla u)^T)$ and $\Omega = \frac{1}{2}(\nabla u - (\nabla u)^T)$). $Q$ represents the local balance between strain rate and vorticity magnitude. A rotational structure is present when $Q > 0$. Physically, it indicates that rotation dominates over strain.

The internal flow pattern differed as we changed the ratio between droplet size and channel width to induce two or three rows of droplets (Fig. 2c-f; see [66] for details of two rows of droplets). The droplets adjacent to the wall had a single rotation. These by-the-wall droplets had either a clockwise or counterclockwise rotation, depending on whether they were by the bottom or top wall. All droplets away from the wall sandwiched by other rows of drops had two counter-rotating vorticity patterns. These mid-row droplets had a clockwise rotation in the upper half and a counterclockwise rotation in the lower half of the droplet. The observed internal flow pattern extends to emulsion flows with 4 and 5 rows of droplets across the width of the channel, respectively (not shown here, see [66] for details).

These observations can be explained by considering the boundary condition at the liquid interface (Fig. 2g). A mismatch in viscosity between the oil and water phases will cause a mismatch in velocity gradient across the water-oil interface. In general, decreasing the viscosity of the disperse phase relative to the continuous phase will increase the velocity gradient in the disperse phase. A large gradient tends to create increasingly pronounced rotational patterns in the droplets [68]. For straight channel flows, the *number* of rotational structures present in each droplet depends on the difference in the velocities of the continuous phase sandwiching the droplet. In turn, the difference in velocities depends on the number of rows of droplets in the channel and the location of droplets within the channel. The *direction* of these rotational structures depends on the ratio of the droplet velocity to the average velocity of the continuous phase sandwiching the droplet. Such a ratio depends on the unique composition of the emulsion (e.g., continuous and disperse phase liquids, surfactants) [69]. Our experiment confirmed that the continuous phase between rows of droplets flowed faster than the droplets (not shown here, see [66] for details), which explains the observed direction of rotation.

We further extended the micro-PIV study to emulsion flows in a tapered microchannel [70]. As mentioned, this geometry is vital because droplets are often interrogated and sorted serially by flowing a concentrated emulsion into a tapered channel with a downstream narrow constriction, whose height and width are comparable to the diameter of a drop [7,20,71]. The tapered geometry imposes a boundary condition that forces the droplets to rearrange via a series of elementary topological processes, also known as T1 events [72]. A T1 event involves the exchange of neighbors among four droplets, where the pair of droplets in initial contact diverges at the end of the T1 event, and the pair of droplets not in initial contact converges at the end of the T1 event (Fig. 3a). From a mechanical point of view, the T1 event corresponds to a transition from one metastable configuration to another after passing through an unstable configuration. This process is driven by the minimization of surface energy [73]. Immediately after a T1 event, strong contact forces move the droplets into their new positions. The relaxation time of the T1 event was found to be determined by the interfacial properties of the droplets [73,74].

T1 events created transient flow fields that had not been observed in a straight channel. In the first step of the rearrangement process, droplet #2 moved diagonally toward the channel wall. Between droplets #1 and #2 (Fig. 3f), a *counterclockwise* vortical structure emerged and then became centered within droplet #1 (Fig. 3g). Between droplet #2 and its downstream neighbor to the right, a *clockwise* vortical structure emerged (Fig. 3f) and then evolved into droplet #2 (Fig. 3g). While droplet #2 was already by the channel wall, it exhibited two vortical structures (a clockwise and a counterclockwise, see Fig. 3g). If there were no T1 events, we would expect a single *counterclockwise* structure in a droplet by the wall. Nevertheless, the *clockwise* structure in droplet #2 (Fig. 3h) weakened over time. Eventually, only a single counterclockwise structure was left. While not shown here, in general, the flow patterns in droplets not participating in T1 events were similar to that in a straight channel.

In the second step of the process, droplets #1 and #3 converged. We observed another pair of counter-rotating structures, one between droplets #1 and #4, and one between droplets #3 and #4 (Fig. 3g). The converging motion of droplets #1 and #3 created these vortical structures. The structures eventually weakened and evolved into flow patterns previously observed in a straight channel without a T1 event. The weakening structures outside the rearrangement zones indicated that these T1-induced flow patterns were transient. Although some of these structures appear centered between two droplets, there was no flow across the droplets.

The flow structures induced by a T1 event increased the circulation inside droplets up to 2.5 times. We calculate circulation within the droplets as the T1 event progressed (Fig. 3i-j). The circulation is related to the strength of a vortical structure, and it can also indicate the degree of mixing inside each droplet. We also observed that the time scale associated with the increase in circulation was approximately the same as the time scale of the T1 event. This observation implies that the changes in circulation were transient, and the droplet motion induced the instantaneous flow pattern inside the droplets during the T1 process. Immediately before or after the T1 event, the differences in velocity between the continuous and the disperse phases set the flow pattern.

Our results imply that the degree of mixing could differ in emulsion droplets, depending on the droplet size relative to the channel size and the position in the channel. If uniform mixing is desired, strategies to swap the droplet position might be necessary. In addition, our results indicate that the geometry of the channel can induce an increase in mixing inside these droplets. For example, to enhance mixing, our results suggest that one can introduce simple variations in the channel width to induce T1 rearrangements.

## 3. Unexpected order in the flow of a two-dimensional concentrated emulsion confined in a tapered microfluidic channel in the low *Ca* regime (*Ca* ~ $10^{-4}$)

When a concentrated emulsion was injected into a tapered microchannel, the droplets self-arranged into a hexagonally packed crystal at static conditions (Fig. 4a). When a slow flow was applied (Reynolds number *Re* ~ $10^{-1}$ and *Ca*~$10^{-4}$ measured at the constriction), we observed that the droplets seemed to always rearrange and "slip" past each other at fixed locations in the channel, which we refer to as "rearrangement zones" (Fig. 4b). The rearrangement zones corresponded to the locations where the number of rows of drops decreased from N to N-1 and were approximately equally spaced. When we examine the instantaneous velocities of individual droplets, we found that they revealed unexpected periodicity in both space and time. A kymograph, in which the magnitude of instantaneous droplet velocity at different *y*-positions in a rearrangement zone (*N* = 7 to 6) is plotted as a function of time, indicates that there was always a drop that moved faster than the others within the rearrangement zone (Fig. 4e-f). The *y*-position of the fast-moving droplet shuttled periodically between the top and bottom walls. The same trend was observed in the kymograph in the non-rearrangement zone (*N* = 7), except it was the position of the slow-moving drop that oscillated between the walls. Similar behavior occurred in other regions of the channel. Furthermore, the directions of the instantaneous velocity vectors of the drops were counter-intuitive from a fluid mechanics point of view (Fig. 4g). For a simple liquid flowing in a converging channel, the velocity vectors should point downward in the upper half of the channel, and upward in the lower half of the channel. Here we observed velocity vectors that pointed upward in the upper half of the channel and vice versa.

Close examination of droplet dynamics shows that the anomalous velocity profiles originated from the rearrangement zones and resulted from a cascade of T1 events (Fig 5a). Within each rearrangement zone, there was one T1 occurring at a time. The T1 propagated along a single direction (+60º or -60º relative to the *x*-axis) until it reached the wall, where it became "reflected" and propagated towards the opposing wall (along the -60º or +60º direction). This shuttling of T1 between the upper and lower walls repeated itself at a nearly constant velocity in a remarkably regular manner not reported before. The periodic patterns in the kymographs resulted directly from the periodic T1 motions: we observed that the diverging droplet downstream (droplet #4) always moved faster than the others. This fast-moving droplet retarded the droplets immediately upstream. As the retardation propagated to the non-rearrangement zone upstream, a similar periodic pattern mirroring that in the rearrangement zone was formed.

Recognizing that concentrated emulsions (and bubble rafts) have long been used as models of crystals for studying grain boundaries, dislocations, plasticity, and other processes central to solid mechanics [43–46,75–77], we decided to attempt to explain these unexpected and highly ordered droplet dynamics by borrowing concepts from solid mechanics. We proceed by treating the concentrated emulsion as a 2D soft crystal and by applying the mechanics of crystal dislocations. The tapered geometry imposed a gradual, elastic compression on the crystal in the transverse direction as it moved along the channel. The T1 cascade can then be considered equivalent to the glide motion of a crystal dislocation on its slip plane. The location of the T1 corresponds to that of a crystal dislocation. The Burgers vector of dislocation can be determined from the Burgers circuit [78] and is at 60º from the *x*-axis (Fig. 5b-c). The dislocation can glide on its slip plane, which contains the Burgers vector. Dislocation motion is driven by the resolved shear stress on the slip plane and is caused by the more significant compressive stress in the *y*- direction than in the *x*-direction [78]. The directions of motion relative to the crystal revealed two types of dislocations in our system (Fig. 5 b-c). In both cases, the *x*-component of the velocities was negative. This negative velocity in *x* canceled the motion of the entire crystal (extrusion) in the positive *x*-direction. Therefore, the dislocation was almost stationary in the *x*-direction relative to the lab coordinate system and only appeared to be shuttling up and down (in the *y*-direction), leading to a localized rearrangement zone.

The periodicity of the rearrangement zones along the *x*-direction can be explained in terms of the Read-Shockley model of low-angle grain boundary [79]. The tapered channel geometry caused a small misorientation ($\theta = 10°$) between the crystal near the top wall and that near the bottom wall. A geometrically necessary dislocation (GND) array was required to accommodate this misorientation. According to the Read-Shockley model, the misorientation here requires a dislocation array with Burgers vector $\boldsymbol{b} = (b_x = 0, b_y)$ and periodicity *s*, such that: $\frac{b_y}{2s} = tan\left(\frac{\theta}{2}\right)$. The Read-Shockley model predicts the periodicity of the rearrangement zones along the channel to be $s = a\sqrt{3}/(4\tan(\theta/2)) = 4.9a$ (for $\theta/2 = 5°$), which is consistent with our observation (Fig. 5d, *s* between the dislocations was ~5.5 droplet diameters).

The most remarkable feature of the droplet dynamics here is the orderly, periodic motion of the dislocations in each rearrangement zone (Fig. 5d-g). To the best of our knowledge, such ordered behavior has not been reported before. Each time a dislocation reached the channel wall and escaped the crystal, a new dislocation was nucleated. The *x*-component of the Burgers vector and the *y*-component of the velocity vector of the new dislocation became opposite to the previous dislocation. This dislocation then traveled along the slip plane with a constant velocity as required by the Read-Shockley model. The whole process subsequently repeated itself. Given dislocation nucleation is usually a stochastic event, these predictable, repeatable nucleation events appeared surprising. The ordered nucleation events occurred here because of the confined geometry of our system: the tapered channel wall applied a compressive load on the extruding crystal, pressing the surface step which formed whenever the edge dislocation reached the channel wall and escaped the crystal. The surface step acted as a local stress concentrator and caused the next dislocation to nucleate at the same location spontaneously. This effect is analogous to heterogeneous dislocation nucleation at a surface step beneath an indenter [80]. Because of this repeated heterogeneous nucleation of dislocations at transient surface steps, the sum of dislocations and surface steps was conserved in the crystal being extruded through the channel. In the absence of dislocation interactions, the sequence of dislocation nucleation and glide events could repeat itself indefinitely with fixed temporal periodicities.

Based on the above discussions, the temporal period $T$ of the dislocation motions can be predicted from the dislocation velocity. By considering the volumetric flow rate and the dislocation velocities, we derived the expected period as $T = \left(N - \frac{1}{2}\right)^2 \frac{\sqrt{3}a^2 H}{2Q}$. Our results show that $T$ is proportional to $\left(N - \frac{1}{2}\right)^2$ (Fig. 5g). The slope of the linear fit is 8.1 ms, very close to the expected value of $\frac{\sqrt{3}a^2 H}{2Q} = 7.6$ ms based on our experimental parameters.

In summary, we have reported the first observation of ordered dislocation dynamics as a two-dimensional soft crystal extruded in a microfluidic channel. While bubble rafts and concentrated emulsions were used as models for crystals, no prior work has described such highly ordered dislocation nucleation and propagation that occurred periodically. This discovery is potentially significant as a model for nanoscale extrusion of solids and flow control in droplet microfluidic systems. As to why such highly ordered flow has not been reported before, it is perhaps due to the requirement to satisfy multiple conditions. The first set of requirements is related to geometric confinement. Both the taper angle and the width of the microchannels must be sufficiently small to ensure the dislocation dynamics prescribed by the Read-Shockley model. The second set of requirements is about defects. The system can only tolerate a low density of defects before the ordered dynamics is disrupted. The system requires monodisperse droplets and a volume fraction of ~85% so that the droplets maintain hexagonal packing. Finally, the flow speed of the emulsion droplets plays a significant role. Upon increasing the flow rate, we found the order was lost. As discussed in the next section, this loss of order from a slow to a fast flow regime is directly associated with the emulsion transitioning from a solid-like to a liquid-like state.

## 4. Timescale and spatial distribution of local plastic events (T1 events): transitioning from low *Ca* to moderate *Ca* flow

The previous section reported periodic droplet rearrangements (T1 events) both in space and time, giving rise to an ordered flow pattern. However, at high flow rates, the order was lost. As the ordered flow pattern in our system arises from elemental T1 events, we hypothesized and showed that the timescale and spatial distribution of the T1 events governed the transition from order to disorder.

We used Voronoi tessellation to identify T1 events (see details in reference [81]). Briefly, we searched for droplets that have five neighbor droplets. At a volume fraction of φ ~ 85%, the droplets were often hexagonally packed in their static configuration, and each drop had six neighbors. During a T1 event, however, one participating droplet (Fig. 6a-b) was surrounded by five neighbors. We extracted the duration of a single T1 event, $T$, by monitoring the evolution of the growing Voronoi cell edges $l(t)$. Depending on the capillary number, $Ca$, the growing edge evolves at different rates. We defined $T$ to be the time required for the edge length to reach 90% of the maximum length and the corresponding non-dimensionalized time scale to be $T' = \frac{T}{\mu R/\sigma}T$.

The dependency of $T'$ on $Ca$ is shown in Fig. 6c. $T'$ can be grouped into three distinct regimes. In Regime 1 ($10^{-7} < Ca < 10^{-6}$), $T'$ was not sensitive to the variation in $Ca$. The scaling followed $T' \sim Ca^{-0.14}$. In Regime 2 ($10^{-6} < Ca < 10^{-3}$), $T'$ decreased with $Ca$. The scaling followed $T' \sim Ca^{-0.61}$. In Regime 3 ($10^{-3} < Ca < 10^{-2}$), $T'$ decreased with $Ca$. The scaling followed $T' \sim Ca^{-0.37}$.

Our results on T1 duration indicate three distinct regimes separated by two transitions at $Ca_{1-2} \sim 10^{-6}$ and $Ca_{2-3} \sim 10^{-3}$ (the subscript denotes the corresponding regimes where the transition occurs). The first transition at $Ca_{1-2} \sim 10^{-6}$ can be explained through scaling analysis of a simplified contact force model that describes the motion of the droplet during a T1 event. We modeled a repulsive force $F_r$ between the drops using the modified Hertz theory for compressed micro-elastomeric spheres with large deformation and approximated frictional force $F_f$ to be negligible due to the small $Ca$ [82,83]. For an emulsion with a volume fraction of $\varphi \sim 85\%$, we obtained a scaling of $T' \sim 10^4 - 10^5$, which is on the same order of magnitude as the lower end of the error bar in our results in Regime 1 ($T' \sim 10^5$).

In Regime 2, as $Ca$ increases, $F_f$ also increases and becomes comparable with $F_r$. Following our scaling argument, we obtain the scaling $T' \sim Ca^{-n}$. Previous studies on quasi-two-dimensional foams and emulsions reported the value of $n$ to be between 0.5 and 2/3 [84–86], depending on the detailed interfacial dynamics [52,57]. Our results in Regime 2 show $T' \sim Ca^{-0.61}$, which lies in the range of the prediction. In our system, the actual interface was likely partially mobile, which could possibly explain why our results in Regime 2 lie between the scaling of $Ca^{-0.5}$ and $Ca^{-2/3}$.

In Regime 3, the drop contact facets are rapidly squeezed, leading to a strong film drainage effect. As a result, the simplified force model for Regimes 1 and 2 no longer applies in Regime 3. To explain the scaling in this regime will likely require a non-equilibrium, time-dependent model. In addition, we observed that in this regime, the droplets in a T1 event did not have sufficient time to relax or complete the initial T1 before the next T1 event started. In other words, in Regime 3, the T1s are coupled and no longer isolated from each other like those in Regimes 1 and 2.

The transition from Regime 2 to Regime 3 coincides with a change in the emulsion velocity profile indicative of a solid-like to liquid-like transition. Below $Ca_{2-3}$, the time-averaged droplet velocity profiles are plug-like (Fig. 7a-b, Fig. 4c-d). These plug-like velocity profiles resemble those described in previous studies, in which emulsions display solid-body motion with zero velocity gradient transverse to the shear direction when the applied strain rate is low [87,88]. This solid-body motion with a plug-like velocity profile arises from the wall slip effect and is considered characteristic of a solid material [64,88]. Above $Ca_{2-3}$, the droplet velocity profiles become increasingly parabolic (Fig. 7c), resembling the profile of the channel flow of a viscous liquid.

Accompanied by the transition from Regime 2 to Regime 3, the droplet trajectories, the spatial distribution of T1 events, and the orientation of successive T1 events also become increasingly different. In Regimes 1 and 2, we observed highly regular and repeatable droplet trajectories. The solid-like behavior is evidenced by the fact that successive T1s occurred along 60° relative to the *x*-axis only (Fig. 7d-e). As detailed in the previous section, the motion of the T1s can be explained purely from a solid mechanics point of view by considering the emulsion as a two-dimensional hexagonally packed crystal and the T1 as a dislocation. In Regime 3, the droplet trajectories occupied the whole channel region without following any specific pattern. The T1 events occurred along the direction of the imposed flow in addition to the 60° direction (Fig. 7f), which is a result of increased wall frictional stress and viscous effects among the droplets [89,90]. The system essentially became liquid-like at high *Ca*.

The change from Regime 2 to Regime 3 as a solid-like to liquid-like transition is further supported by plotting the results plotted against the Deborah number $De = \frac{T_r}{R/U}$, where $T_r$ is the relaxation timescale of the flow, and $R/U$ represents the timescale associated with the droplet advection (Fig. 6d). In terms of dimensionless quantities, $De = T_r' * Ca$, where $T_r'$ is the dimensionless relaxation timescale. $De$ delineates solid-like versus liquid-like regimes. If $De < 1$, the drops have sufficient time to complete the full relaxation process during the droplet advection, and the material is solid-like. On the other hand, if $De > 1$, the drops cannot fully relax before they participate in the subsequent rearrangement, and the material is liquid-like. The crossover between the two regimes can be obtained by setting $De = 1$. Setting $De = 1$ gives $Ca_{S-L} \sim 2 * 10^{-3}$, close to the transition between Regime 2 and Regime 3 at $Ca_{2-3} \sim 10^{-3}$. Therefore, we can conclude the transition from Regime 2 to Regime 3 likely corresponds to a transition of the emulsion from a solid-like state to a liquid-like state.

Additionally, the main difference between Regime 1 and Regime 2 lies in the single T1 event duration $T'$: in Regime 1, $T'$ is insensitive to $Ca$; in Regime 2, $T'$ decreases with $Ca$. Since $T'$ in these two regimes is not the rate-limiting factor and does not enter the calculation of $De$, it explains why the flow features (droplet trajectories and velocity profiles) are similar in Regime 1 and Regime 2.

Finally, we relate the above discussion to the order-to-disorder transition in the microfluidic crystal. Previously, we identified that the ordered flow patterns of microfluidic crystal were lost at $Ca \sim 10^{-2}$ measured at the constriction [90]. This value of $Ca$ corresponds to the beginning of Regime 3 in the present study. Below this value, the emulsion was in Regimes 1 and 2 and was expected to be solid-like. The T1s were isolated and did not interact. Their propagation was analogous to a dislocation glide along the crystal plane, and the resulting flow profile was highly ordered. As the applied flow rate and the corresponding $Ca$ increased above $Ca_{2-3}$, the emulsion transitioned to Regime 3 and was expected to become liquid-like. In this regime, successive T1 events occurred along the imposed flow direction. These T1s interacted with the T1s along the 60º slip plane. Such interactions could lead to various outcomes (e.g., the mutual blockage or annihilation of T1 dislocations), disrupting periodic slip dynamics and ordered flow behavior. In other words, the loss of order in the flow of the microfluidic crystal as the flow rate increases originates from the emulsion transitioning from a solid-like material to a liquid-like material.

## 5. Breakup of concentrated emulsion droplets in the high $Ca$ regime ($Ca > 10^{-2}$)

In the previous sections, we reviewed some intriguing behaviors of emulsion flows driven at low to moderate $Ca$. In a low-rate regime (capillary number $Ca \sim 10^{-4}$) where interfacial effects dominate viscous effects, the emulsion flow could be modeled as a solid crystal. When the flow rate was increased to a medium rate regime ($Ca \sim 10^{-3}$), the interfacial and the viscous effects became equally important. As a result, the emulsion flow experienced a solid-like to liquid-like transition. This section centers on a high-rate regime ($Ca > 10^{-2}$), where the viscous effect starts to challenge the stability of the droplet liquid-liquid interface. In this high-rate regime, droplets can undergo breakup, compromising the accuracy of the assay. The undesired breakups remain a crucial bottleneck limiting the assay throughput [7,20]. Although sufficient for some current applications, this limitation will restrict droplet microfluidics for applications that require a further increase in throughput. An ideal solution would involve effective methods to prevent the breakup. We first discuss the process and some critical parameters that delineate the breakup of emulsion droplets in microchannels. Second, we will present several practical strategies to mitigate the undesired breakup. We focus the discussion on droplet breakup in a tapered microchannel with a narrow constriction that forces droplets to pass through the interrogation region one at a time. As mentioned in the introduction, this channel design is widely adopted in high-throughput applications for serial interrogation, sometimes followed by sorting.

To characterize droplet breakup in our system, we defined the droplet breakup fraction as the number of droplets that broke up divided by the total number of droplets that flowed through the microchannel. We extracted the number of breakup events (Fig 8a-b) by comparing droplet size distributions upstream and downstream.

In the tapered microchannel, we note that a single drop did not break at the fastest flow condition tested [20]. Breakup, therefore, arises from droplet-droplet interactions, often when multiple droplets attempt to enter the constriction simultaneously, and one drop pinches another drop against the channel wall. We investigated the breakup fraction as a function of droplet size, constriction geometry, viscosity ratio, and flow rates (Fig. 8c-d). The breakup fraction increased with the capillary number for a fixed set of droplet size, constriction geometry, and viscosity ratio. Physically, an increasing capillary number corresponds to increasing viscous stress experienced by a drop, which becomes increasingly deformed. In our system, more deformed drops were also more prone to breakup than less deformed drops [20]. However, plotting breakup fraction vs. *Ca* did not collapse the data to a single curve. This result indicates that, unlike the breakup of single droplets in an unconfined flow, the capillary number alone was insufficient to describe the breakup phenomena completely. From dimensional analysis, it is possible to identify additional governing parameters to consist of the confinement factor ($cf = \frac{r}{r_h}$, where $r$ is drop radius calculated by assuming the drop is spherical, and $r_h \equiv \frac{W_c H}{(W_c + H)}$ is the hydraulic radius of the constriction with $W_c$ and $H$ being the width and height of the constriction, respectively) and viscosity ratio ($\lambda = \frac{\mu_d}{\mu_c}$, where the subscript *d* denotes the disperse phase and *c* denotes the continuous phase). Indeed, when the breakup fraction was plotted against *Ca\*cf\*λ*, the data collapsed nicely into one single curve (Fig. 8e).

A key difference between the breakup of droplets inside a concentrated emulsion from the breakup of a single droplet is that the former is not deterministic. The next question we ask is what gives rise to the probability distribution? We examined droplets close to the constriction, where most breakup events occurred (Fig. 9). We observed that most breakup events occurred between two drops pinching each other when they entered the constriction. The fate of the drops—whether breakup occurs or not—depends strongly on the relative position between the two drops as they enter the constriction. We measured the offset between the leading edges of the two closest drops entering the constriction. We found that there exists a critical offset ($\Delta x_{cr\_1}$) below which breakup always occurs ("Region 1"), and a critical offset ($\Delta x_{cr\_2}$) above which no breakup occurs ("Region III") (Fig. 9). There is a narrow bistable region between these two critical offsets where both breakup and non-break-up events exist ("Region II") [91,92]. We believe that in the bistable region, the contribution from additional droplets determines whether breakup occurs or not.

Overall, this finding suggests that the probability of droplet breakup in a concentrated emulsion is governed by, at least in part, the probability that two drops are synchronized in their entry into the constriction (i.e., having a small offset between their leading edges). The result in Fig. 8e has immediate implications for guiding the channel design to prevent undesired droplet breakup while preserving the interrogation rate. For example, if the sample contains protein or other polymer solutions that has double the viscosity of water (i.e., $\lambda = 2$), one should then use a constriction geometry with half of the confinement factor to maintain the same breakup fraction (ideally a low fraction) while operating at the same capillary number or droplet interrogation rate [30].

## 6. Strategies to mitigate droplet breakup within a concentrated emulsion

With the breakup process and key parameters identified, a natural question is: how can we suppress these undesired droplet breakups? This section summarizes some practical strategies in our recent work to suppress droplet breakup.

### 6.1 Using nanoparticles to stabilize emulsion droplets

Some prior work demonstrated that tuning the droplet interfacial viscoelasticity impacted droplet deformation. For example, a particle-stabilized drop deforms less than a pure liquid drop at comparable flow conditions [93]. Given the close correlation between droplet deformation and breakup in our system, we hypothesized that droplets stabilized by nanoparticles (NPs) would have a lower breakup fraction than droplets stabilized by surfactants at the same flow condition. Another motivation for replacing surfactants with NPs was that NPs mitigate surfactant-mediated inter-drop transport of small molecules such as drugs and fluorophores [8,9,30,94]. Such transport leads to the undesirable cross-talk of droplet contents and destroys the accuracy of droplet-based assays [95]. In addition, we found that NPs provide a sufficiently rigid interface that supports the attachment and growth of adherent cells, a capability not possible in surfactant systems.

To test the hypothesis, we synthesized amphiphilic fluorinated silica nanoparticles [9] to stabilize droplets and generate emulsions with a volume fraction consistent with all previous experiments (volume fraction $\varphi \sim 85\%$). We found the droplet breakup fraction was significantly lower than that of surfactant drops at the same values of $Ca*cf$ (Fig. 10a) [10]. In addition, the breakup fraction of NPs-stabilized drops was not sensitive to $\lambda$ within the range tested (Fig. 10b). These results indicate two critical advantages of NPs with direct applications in droplet microfluidics. First, since droplet breakup is insensitive to viscosity ratio, one can use the same droplet size and channel geometry (i.e., the same confinement factor) for samples having different viscosities (e.g., polymer or protein solutions that can be more viscous than water) and maintain the same degree of breakup and corresponding assay accuracy. This feature is an advantage over the surfactant system studied (Fig. 8e). Second, replacing surfactants with NPs increases the throughput of the serial droplet interrogation process. For example, suppose 3% of droplet breakup is acceptable in a microfluidic assay. In that case, the highest throughput is ~7100 drops/second for 70 pL, NP-stabilized droplets in a 30º taper with a 1.51 confinement factor. This throughput is three times higher than surfactant-stabilized droplets studied here.

### 6.2 Reducing the volume fraction of emulsions

For droplets within a concentrated emulsion (an emulsion with a high disperse-phase volume fraction of $\varphi > 0.80$), we have discussed that their breakup arose from droplet-droplet and droplet-wall interactions [10,20,71,91]. Previously, we found that a single drop did not break at the fastest flow condition we have tested [20]. Such observation indicates that the volume fraction of the emulsion plays a critical role in droplet breakup.

To elucidate the relationship between droplet breakup and the volume fraction of the emulsion, we diluted the concentrated emulsion on-chip immediately upstream of the entrance of the constriction (Fig. 11a-b) [96]. As expected, the breakup fraction decreased with decreasing effective volume fraction of the emulsion (Fig. 11c-d). The breakup fraction increased significantly with an increasing capillary number for $\varphi > 0.60$, while it increased slightly for $\varphi < 0.50$ (Fig. 11d) [96]. These results are consistent with our previous observations that breakup arises primarily from droplet-droplet interactions when multiple droplets attempt to enter the constriction simultaneously, and one drop pinches another drop against the channel wall. The frequency of occurrence of such interactions is expected to decrease significantly below the random close packing limit of the emulsion as the average inter-drop distance increases. These results indicate that while a higher emulsion volume fraction packs more drops per unit volume, the propensity of the drops to undergo breakup limits droplet throughput if droplet integrity and assay accuracy are to be maintained. For example, at a droplet breakup fraction of 0.10, diluting the emulsion 2.1 times from $\varphi = 0.85$ to $\varphi = 0.40$ increases the droplet throughput by ~1.5 times (Fig. 11e) [96].

**6.3 Strategic placement of an obstacle upstream of the constriction**

When granular materials, colloidal suspensions, and even animals and crowds exit through a narrow outlet, clogs can form spontaneously when multiple particles or entities attempt to exit simultaneously, obstructing the outlet and ultimately halting the flow [97–101]. Counterintuitively, the presence of an obstacle upstream of the outlet has been found to suppress clog formation in the flow of rigid particles [97,102,103] and pedestrians evacuating a room [104–106]. Analogous to how an obstacle reduces clogging in these systems, we hypothesize and demonstrate that an obstacle could suppress breakup in our concentrated emulsion flowing in a tapered microchannel (i.e., a 2D hopper) by preventing the simultaneous exit of multiple drops.

We tested this hypothesis by investigating droplet breakup fraction $\beta$, in a 2D hopper with different obstacle sizes and positions (Fig. 12a-b) [107]. Overall, the obstacle modified the flow of the droplets in its vicinity to prevent droplet breakup. Remarkably, even though the obstacle introduced narrow gaps with the sidewalls, which would typically increase the shear stress the drops experienced and make them more prone to breakup, the breakup fraction did not increase for most values of the obstacle-to-constriction distance, $x$. On the contrary, as obstacle distance $x$ decreased, the breakup fraction decreased significantly by almost $10^3$-fold and then increased rapidly as the obstacle approached the constriction.

To probe the origin of breakup suppression by the obstacle, we examined the packing of the drops around the obstacle. We observed a common pattern when minimum breakup probability occurred: (i) only one drop occupied the gap between the obstacle and the sidewall at a time (see the red drops in Fig. 12d region V);(ii) only two drops occupied the whole width of the channel immediately downstream of the obstacle (see the green drops in Fig. 12d region V). This pattern was further supported by plotting all experimental data in a regime map of $l_{ow}/D_d$ against $l_{pow}/D_d$ (Fig. 12c), where $l_{ow}$ is the minimum distance from the obstacle to the sidewall, and $l_{pow}$ is the minimum distance from the posterior end of the obstacle (closest to the constriction) to the sidewall. Across all obstacle geometries, droplet sizes, and flow rates tested, droplet breakup was effectively suppressed when both $l_{ow}/D_d$ and $l_{pow}/D_d$ approached one.

The strategic placement of an obstacle in the hopper induces passive ordering of the drops in the vicinity of the obstacle. When optimally placed, we found that the obstacle facilitates a hexagonal or zig-zag packing downstream of the obstacle. This packing causes the drops to alternate and enter the constriction in a single file in order (Fig. 12e). Accordingly, the streamwise offset between the leading edges of two consecutive drops entering the constriction was larger in region V than in all other regions. As discussed previously, breakup arises from the simultaneous entry of two drops into a constriction [91,92]. Our results suggest that an optimally-positioned obstacle can mediate the ordering of the drops, which prevents simultaneous entry into the constriction and subsequent breakup [107].

In summary, introducing an obstacle can offer a passive and straightforward strategy to minimize breakup by almost three orders of magnitude than when the obstacle is absent. It can be used to increase the throughput of the droplet interrogation process and improve the robustness of droplet-based biochemical assays.

## 7. Computational methods to model crowded drops in microfluidics

A thorough understanding of the unique fluid dynamics features revealed by droplets in microfluidic environments requires a reliable and accurate mathematical modeling of fluid interfaces to capture the multiscale physics of the phenomena in play.

From a methodological standpoint, bridging the scales between the interface and the device is not an easy task due to the different nature of the complex interactions involved. These complex interactions have been the subject of the seminal works of Derjaguin, Landau, Verwey, and Overbeek, which culminated in the so-called DLVO theory [108,109]. The scale separation between DLVO interactions and the device operating length extends over five to six decades. Indeed, interaction forces act at scales as small as nanometers, hydrodynamic relevant phenomena emerge three decades above, and the typical length scale of the device reaches the centimeter scale.

From the above, the direct introduction of interfacial forces at a molecular level reflects the need to solve simultaneously six spatial decades, a task which is simply out of reach to date [110].

On the other hand, [111], one can think of upscaling the effects of the relevant forces occurring at the interface level by defining suitable pseudo-potentials containing enough information to reproduce (at least) the phenomena occurring at the droplet and the device scales. In the absence of such a universal behavior, one cannot escape the use of direct molecular approaches, making it impossible to reach up to the size of the full device.

We now try to briefly summarize the efforts made so far in the computational physics of fluid interfaces.

When dealing with the physics of fluid interfaces, it is possible to distinguish between two main approaches, i.e., sharp and diffuse interfaces. In the sharp interface description [112], introduced by Young, Laplace, and Gauss back in the 19$^{th}$ century, the fluid interface is considered as a discontinuity between two immiscible phases. The physical quantities such as density and viscosity are also assumed discontinuous across the interface.

On the other hand, at the end of the 1800s, Van der Waals and Lord Rayleigh came out with the idea of a diffuse interface, namely a smooth transition between two immiscible fluids [113]. The idea was then further investigated by Korteweg, who proposed a constitutive law for the capillary stress tensor defined in terms of spatial gradients of suitable order parameters [113].

The same distinction can be identified in computational fluid dynamics.

In both diffuse and sharp interface models, the dynamics of the bulk fluids is described by a set of coupled, partial differential equations for mass and momentum conservation, namely the Navier-Stokes equations:

$$\partial_t(\rho\boldsymbol{u}) = -\nabla \cdot (p\boldsymbol{I} + \rho\boldsymbol{u}\boldsymbol{u} + \boldsymbol{T} - 2\mu\boldsymbol{D})$$

where $\boldsymbol{u}$ is the velocity vector, $p$ is the pressure, $\rho$ is the fluid density, $D$ is the rate of strain tensor multiplied by the dynamics viscosity $\mu$, and $T$ is the capillary stress tensor.

The extended pressure tensor accounts for the necessary information needed to track the complex evolution of the interface dynamics.

**7.1 Sharp interface models**

In sharp interface approaches, the interface is described as a moving boundary across which the surface tension at the fluid-fluid interface is obtained by imposing a stress balance needed to reproduce a Laplace-like pressure jump at the interface:

$$(-p\boldsymbol{I} + 2\mu\boldsymbol{D}) \cdot \boldsymbol{n} = 2\sigma k\boldsymbol{n}$$

In the equation above, σ is the interfacial tension, $k$ is the local curvature of the interface, and **n** is normal to the interface.

Marker methods and Volume of fluids fall within this modeling philosophy [114,115]. In the former, tracers or marker particles are used to locate the phases. In two-phase flows, the velocity field is advanced by solving the Navier-Stokes equations on an Eulerian grid. The interface is then evolved by advecting each marker due to the action of the local fluid linear momentum. A high degree of accuracy may be obtained in Marker methods due to the possibility of employing a high-order interpolation polynomial to represent the interface [116].

A particular case of the marker approach is represented by the boundary integral method (BIM) [117,118]. In BIM, the evolution of a fluid interface, deforming and moving in space and time, is obtained via time integration of the fluid velocity of a set of marker points positioned at the interface. The velocities of the marker points are obtained by solving a boundary integral equation, and the flow solution is inferred from the information of the discrete points along the interface. In the field of droplet microfluidics, BIM has been extensively employed in studying the deformation of confined droplets, the dynamic evolution of surfactant-laden droplets in three dimensions, and, more recently, the deformation and breakup of droplets flowing in T-junction and concentrated emulsions through narrow constrictions [92,119–122].

The volume of fluid methods makes use of a suitably defined function aimed at identifying the volume fraction of each fluid in each grid cell. The interface is then advected and reconstructed at each time step. Several approaches may be exploited for the reconstruction of the interface such as stepped approximations, piecewise constant, or a higher-order approximation such as PLIC (piecewise linear interface construction) or spline [112]. The VOF approach has been recently employed to investigate droplet formation in T-junction devices [123] and to predict the dripping to jetting transition and the contact angle for which droplets form in a step emulsifier device [124].

Recently, a VOF code has been implemented to deal with systems with non-coalescing interfaces [125]. The approach is based on a standard finite volume discretization based on Chorin's projection method, in which the advection equation is solved using the VOF method with piecewise linear PLIC reconstruction. To overcome the coalescence between neighboring interfaces, the multimarker volume-of-fluid method is employed [126], which completely suppresses coalescence between neighboring interfaces. The drawback of this approach lies in the impossibility of tuning the repulsive interfacial forces due to the presence of amphiphilic surfactants or colloids adsorbed at the fluid-fluid interface.

**7.2 Diffuse interface models**

The assumption of a sharp interface breaks down when the interfacial thickness is comparable to the length scale of the phenomena under scrutiny. For example, investigating the motion of a contact line along a solid surface may require setting the length scales comparable to that of the interface thickness.

From this perspective, diffuse interface models represent a viable alternative to sharp interface models for capillary-driven interfacial phenomena [113].

In diffuse interface models, field quantities are continuously distributed throughout the interfacial region, and surface tension and interface motion emerge naturally by directly including the capillary (or Korteweg) stress tensor in the generalized pressure tensor.

Phase-field models [113] have been extensively employed in the field of multiphase and multicomponent flows for the simulation of many complex interface phenomena (e.g., spinodal decomposition [127], mixing, and interfacial stretching [128], moving of contact lines and homogeneous nucleation [129,130]).

An important contribution to diffuse interface modelling in the last two decades has been brought about by the Lattice Boltzmann method (LBM) (see [131–134] for a detailed description of the LBM).Many extensions of the method have been proposed in the literature to capture the behavior of multiphase and multicomponent systems.

The pseudo-potential approach was one of the first diffuse interface models to be developed in the context of the LBM [135]. This method mimics the fluid interactions via an interparticle potential, through which the separation of different phases or components can be achieved naturally. A second option is represented by the free-energy (FE) LB method developed by Swift et al. [136]. In FE LBM, the pressure tensor is explicitly modified to include a nonideal thermodynamic pressure tensor. An attractive feature of FE LB models is that, by design, they are always thermodynamically consistent, in contrast with the pseudo-potential approach. Another choice is represented by the color-gradient approach, first introduced by Rothman and Gunstensen [137]. In this case, the collision step is a three-step process: a standard collisional relaxation, a perturbation step that contributes to the build-up of the surface tension, and a recoloring step that mimics the segregation between immiscible phases.

The above methods have been extensively employed to simulate extremely complex phenomena at the micron scales, such as spinodal decomposition, droplet collisions, and dense emulsions under shear. Multicomponent Lattice Boltzmann methods can also be coupled to molecular dynamics algorithms developed for suspended (colloidal) particles to study particle-laden interfaces. This approach resolves not only hydrodynamics but also individual particles and their interactions at the interfaces [138–140]. The use of such algorithms led to discovering a new material known as Bijels (bicontinuous interfacially jammed emulsion gels) in early 2000 [141], a preciously rare case in which simulation predicted a new material ahead of experiments.

Notwithstanding the above, from a computational standpoint, the inclusion of repulsive near-contact interactions, needed to prevent fluid-fluid interfaces from coalescence, still represents an issue. Recently, an LBM model for multicomponent fluids augmented with near-contact interaction forces has been developed to deal with the above issue [110,142]. The augmented approach has been employed to predict the formation of soft granular materials in flow focuser devices, capture the self-assembly of droplets in microfluidic channels, and simulate the wet-to-dry transition in very dense emulsions [143]. This approach opened the possibility of simulating advanced microfluidic applications which were not accessible to previous LBM schemes.

**7.3 Examples of computational models on concentrated emulsions**

The computational strategies described above have been extensively employed to capture the complex dynamics underlying the evolution of fluid interfaces across a wide range of spatio-temporal scales. In droplet microfluidics, VOF-based approaches coupled with different numerical schemes for solving mass and momentum equations have been recently applied to simulate the formation of emulsions in a microfluidic device. Aza and co-workers [144] investigated the impact of the main dimensionless numbers governing the formation of double emulsions in a hierarchical flow by employing the VOF-finite volume approach augmented with adaptive mesh refinement (Fig. 13a). More specifically, the authors are interested in assessing the effect of the governing parameters on the droplet generation frequencies, the inner/outer droplet size, and the position of the pinching point. In [123], a PISO (Pressure Implicit with Splitting of Operator) VOF method was used to simulate the droplet detachment in T-junction geometries in a wide range of capillary numbers and viscosity ratios (Fig. 13d). More recently, a high-performance computing software for multiphase flows based on a finite volume discretization based on Chorin's projection method and a multi-color VOF with piecewise linear PLIC [125] reconstruction has been employed to simulate the droplet clustering at the outlet of a divergent channel, showing good agreement with recent experimental evidence (Fig. 13b), as well as the formation of fluidic crystals in microfluidic channels.

Zinchenko et al. showed the capability of a multipole-accelerated 3D boundary-integral algorithm to model the pressure-driven flow of a highly concentrated emulsion of deformable drops through a periodic channel with tight constrictions [119]. These simulations, inspired by experiments reported in [20], aimed at understanding how dense monodisperse emulsions behave as monolayers between two tight, parallel walls when they enter a constriction much narrower than the non-deformed drop diameter (Fig. 13c).

As per diffuse interface approaches, huge efforts in the modeling of dense microfluidics emulsions have been deployed in the field of lattice Boltzmann modeling. Pseudo-potential-based multicomponent LB models have been extensively employed to investigate the dynamics of dense foams in Couette cells [145], in close agreement with experimental evidence as visible in Fig. 14a, and to investigate the solid-to-liquid transition in soft-glassy fluids with an imposed spatially heterogeneous stress [146].

More recently, a class of multicomponent LB models, augmented with repulsive near-contact interactions, have been shown to: i) correctly reproduce the formation of soft-flowing crystals in microfluidic channels [142] in agreement with previous experimental findings (Fig. 14b); ii) simulate the wet-to-dry transition in dense emulsions depending on the capillary number and the device geometry (i.e., angle of aperture of the divergent channel) and the self-assembly of droplets in microfluidic devices [142,143]; iii) capture the deformation and breakup dynamics of droplets within a tapered channel (Fig. 14c) [110]; iv) predict a class of novel dynamical modes in soft granular materials, namely flowing packed double emulsions, in flow focusers (Fig. 14d) [147].

Still in the LB framework, the free-energy approach (FELB) has been used to investigate the dynamics of dense droplet-based systems. As an example, Marenduzzo and co-workers used FELB to study suspensions of soft deformable droplets [148], showing that their rheology undergoes discontinuous shear thinning behavior under a pressure-driven flow (Fig. 14e). The Authors also showed that such a discontinuity may be viewed as a nonequilibrium transition between a hard droplet regime, which flows slowly, and a soft droplet phase, which flows much more readily. A similar numerical approach has been recently used by Tiribocchi et al. [149] to predict the vortex-driven dynamics of droplet in dense double emulsions. In this study, the authors investigate the physics of multi-core emulsions flowing in microfluidic channels reporting numerical evidence of a rich variety of driven non-equilibrium steady states (NESS), whose formation is caused by a dipolar fluid vortex triggered by the sheared structure of the flow carrier within the microchannel.

To conclude, the difficulty of performing a "comprehensive" multiscale droplet microfluidic simulation lies in the fact that these classes of problems span six or more orders of magnitude in space (typically from nm to mm) and nearly twice as many in time. To address such a computational challenge, the next generations of computational scientists will be faced with the task of finding imaginative and efficient solutions combining High-Performance Computing techniques, adaptive grid refinement strategies, coarse-grained models, Machine Learning procedures aiming at preconditioning the numerical solution to accelerate the convergence of very large-scale simulations [150] and, on a longer term perspective, probably quantum computing as well [151,152].

## 8. Conclusions

In summary, our studies in concentrated emulsions have revealed new regimes of droplet order and instability that cannot be predicted by prior work on dilute emulsions. Our understanding of how drops interact has taught us how to engineer novel microfluidic designs to overcome or leverage such interactions to enhance the performance of droplet technology, which in turn lays the foundation for advancing droplet technology for a wide range of applications. Beyond droplet microfluidics, emulsions are important for emerging applications such as 3D printing, modeling of biological tissues (as active emulsions or foams), and biomimetic materials. We anticipate our studies could be applicable towards predicting the behavior of these systems as well.

**Acknowledgement**

This work was supported by NSF Award No. 1454542. We are grateful for insight discussions with Professors Howard Stone, Dave Weitz, John Dabiri, Sascha Hilgenfeldt, Michael Brenner, Wei Cai, Tobias Schneider, Robert Davis, and Claudiu Stan. AM and SS wish to acknowledge financial support from the European Research Council under the Horizon 2020 Programme Grant Agreement n. 739964 ("COPMAT").

Figure 1. Summary of our work on concentrated emulsions.

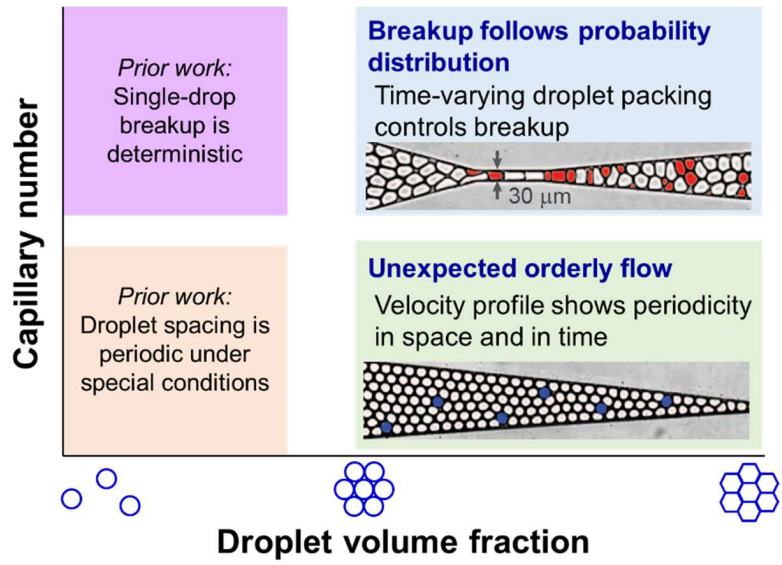

Figure 2. **(a)** μPIV experimental setup for flow measurements of droplets within a concentrated emulsion in a microfluidic channel. **(b)** A representative instantaneous fluorescent raw image was acquired using the μPIV technique. **(c)-(f)** Representative three rows of droplets: **(c)** Instantaneous and **(d)** ensemble-averaged velocity vector fields. The droplet centroid velocity is subtracted from all the velocity vectors shown. The velocity vectors are colored by the magnitude of normalized total velocity, $U_{tot}^*$, which is normalized by the droplet centroid velocity. **(e)** Color contours of normalized vorticity field, $\omega^*$, computed using the ensemble-averaged results. Vorticity is normalized by the droplet diameter and droplet centroid velocity. Red and blue indicate counterclockwise and clockwise vorticity, respectively. **(f)** Color contours of normalized $Q$-criterion, $Q^*$, computed using the ensemble-averaged results. Colors are shown based on the results of vorticity. The spatial coordinates of $x^*$ and $y^*$ are normalized by the width of a single-drop wide channel (=50 μm). **(g)** A sketch illustrating flow inside a single row, two rows, and three rows of droplets within a concentrated emulsion spanning the entire width of the channel. The velocity profiles inside the droplets (black lines) are based on the experimental results, while those in the continuous phase (shown in the insets) are proposed based on the no-slip and matching shear stress boundary conditions at the water-oil interface (grey lines). All velocity profiles shown are relative to the droplet velocity. Red and blue arrows show the counter-clockwise and clockwise vorticity, respectively. Reproduced from [66], with the permission of AIP Publishing.

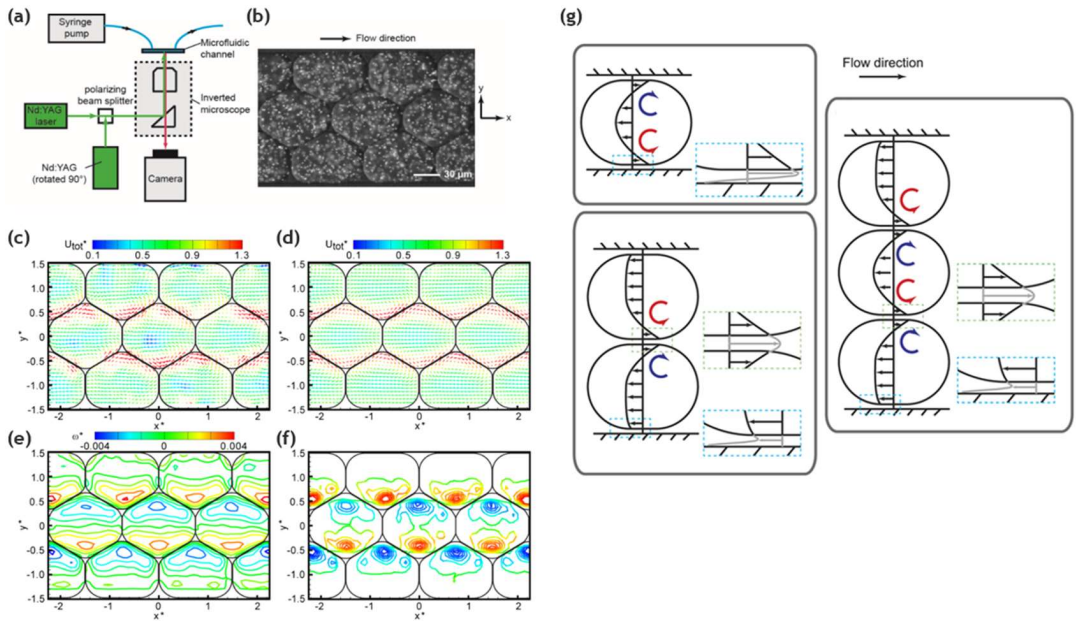

Figure 3. **(a)** T1 droplet rearrangement process. Droplets labeled #1 and #3, which were not in contact initially, converge at the end of the T1 process. Droplets labeled #2 and #4, which were in contact initially, diverge at the end of the T1 process. **(b)** A representative instantaneous fluorescent raw image was acquired using the μPIV technique. **(c)-(h)** Phase-averaged flow field in the rearrangement zone N = 3 to N = 2. **(c)-(e)** Velocity vector field and **(f)-(h)** Q-criterion for three time points within one T1 event at t = 0T, 0.5T, and 1.0T, where T is the period of a T1 event. The T1 event occurs at an angle of +60º with respect to the flow direction. Droplets involved in the T1 event are outlined by solid lines while the ones not involved are outlined by dotted lines. Flow is from left to right. Velocity vector and its color are shown relative to the droplet velocity and the color also represents the phase-averaged normalized total velocity, $\widetilde{U}_{tot}^*$. Phase-averaged normalized Q-criterion, $\widetilde{Q}^*$ is colored by the phase-averaged normalized vorticity, $\widetilde{w}^*$. **(i)** Non-dimensional circulation, $\Gamma^*$, vs. non-dimensional time, $t_{N32}^*$ for the four droplets involved in the rearrangement process. Here, circulation is non-dimensionalized by circulation inside the droplets at the constriction, and time is non-dimensionalized by the duration of the T1 event for this rearrangement zone (i.e., from $N = 3$ to $N = 2$). Every other data point is shown for clarity purposes. Solid lines are curve fits to the data points and are guides for the eyes only. The error in the data points estimated using the propagation of uncertainty analysis is about $\Delta\Gamma^* = \pm0.37$. **(j)** Non-dimensional average velocity inside a droplet, $U_{drop}^*$ vs. non-dimensional time, $t_{N32}^*$ for the four droplets involved in the rearrangement process. Here, the average velocity inside a droplet is non-dimensionalized by the droplet centroid velocity at the constriction. Reproduced from [70], with the permission of AIP Publishing

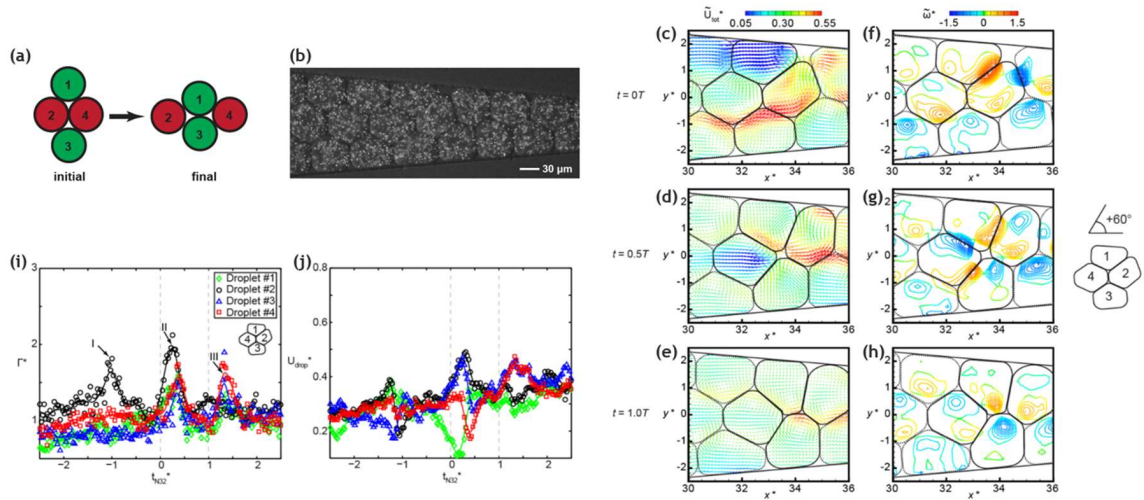

Figure 4. **(a)** Microscope image of the concentrated emulsion in the tapered microfluidic channel. **(b)** The time-averaged total velocity of drops at various locations in the channel. The velocity was normalized to the maximum total velocity measured at the constriction. The position of each marker represents the position where the droplet velocities were averaged. The color of the marker represents the magnitude of droplet velocity at that location. Note that the markers overlap and appear as continuous lines which reflect the trajectories of the drops. The rearrangement zones are indicated by blue arrows. **(c)-(d)** Time-averaged x-component of droplet velocity $\overline{u_x}$ across the width of the channel measured at the locations marked by the blue and green boxes in b. The velocities shown were normalized to $\overline{u_{x,max}}$, the maximum x-component of velocity measured in the constriction. **(e)-(f)** Kymographs of instantaneous droplet velocities in the rearrangement zone (N = 7 to 6), and non-rearrangement zone (N = 7), respectively. N is the number of rows of drops across channel width. The position (t, y) of each marker represents the time t and the y-position of a drop in the channel when the velocity of the drop was measured. The color of the marker represents the magnitude of the total velocity of the drop normalized to the maximum velocity measured at the constriction. **(g)** Instantaneous velocity vectors of the drops in the channel. Reproduced from [90].

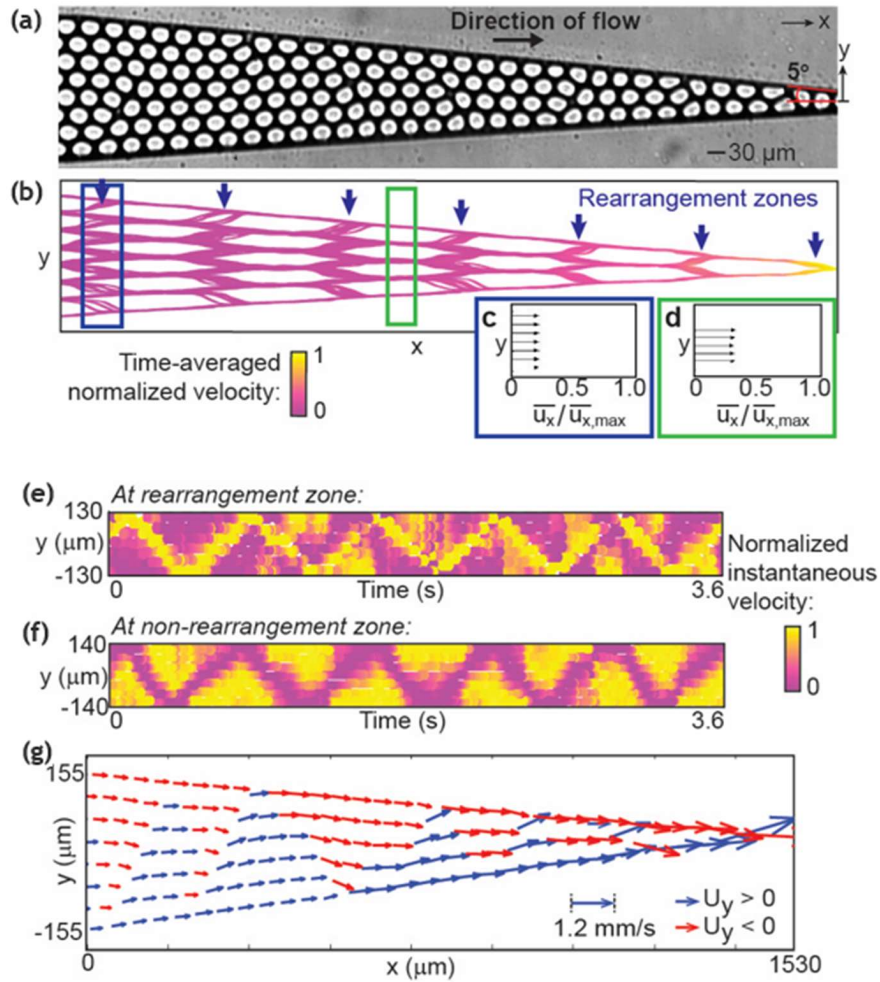

Figure 5. **(a)** A T1 event involves one pair of diverging drops (#2 and #4) and one pair of converging drops (#1 and #3). Snapshots of the T1 cascade which is equivalent to a dislocation glide. Three T1 events were highlighted: event 1 (t = 0 to 0.13 s) where the diverging (converging) drops were colored in red (green); event 2 (t = 0.13 to 0.25 s) where the diverging (converging) drops were colored in purple (yellow); event 3 (t = 0.25 to 0.38 s) where the diverging (converging) drops were colored in red (green). **(b)-(c)** Burgers circuits (pink lines) of the dislocations in the crystal at t = 0.19 s (also indicated by the red box in a), and at t = 0.58 s, respectively. The pink arrow indicates the Burgers vector b. The green line indicates the slip plane, and the green arrow indicates the direction of motion of the dislocation. The black arrows indicate the stress acting on the crystal. **(d)-(g)** Spatial and temporal periodicity of dislocation dynamics: (d) $x/a$-positions of T1s in the channel as a function of time. The mean spacing $s$ between the seven sets of slips shown is $s = 5.5$. $a$ is the diameter of one drop. Each trace represents the $x/a$-positions of the T1s within one rearrangement zone where N rows of drops reduces to N-1. **(e)** Fluctuation in the $x/a$-position of the dislocations as a function of channel width expressed in terms of the number of rows of drops N. The height of the gray error bar represents the maximum fluctuation x/a measured (also plotted in the inset). **(f)** Scaled y-position ($y'$) of the dislocations as a function of time. For each trace, the maximum (or minimum) $y'$ value represents the top (or bottom) of the channel wall within one rearrangement zone where $N$ rows of drops reduces to $N-1$. The $y$ positions of the T1s oscillate with period $T$. For both a and c, the positions plotted are that of the drops with five neighbors for convenience. **(g)** The period $T$ scales linearly with $(N - 1/2)^2$. The periods are calculated using the Fast Fourier Transform of the data in c. Gray markers represent data from four independent experiments at the same flow conditions. Black markers represent the average from these experiments. The black line is the best linear fit to the average. Reproduced from [90].

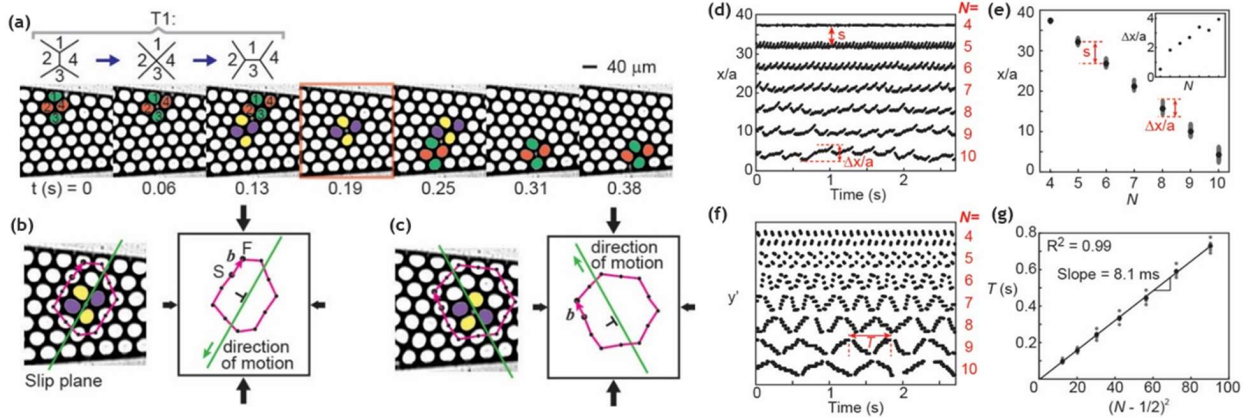

Figure 6. **(a)-(b)** Voronoi tessellation during a T1 event at **(a)** $Ca = 4.9 * 10^{-7}$ and **(b)** $Ca = 2.1 * 10^{-3}$ respectively. The green lines represent the Voronoi cell edges. The red line represents the growing edge shared by converging drops during the T1 events. The blue dots represent the centers of 5-edged Voronoi cells. **(c)** Log-log plot showing $T'$ as a function of $Ca$. Each data point consists of a batch of $T'$ measurements of >200 T1 events within one rearrangement zone at a given flow rate $Q$. The vertical and the horizontal error bars represent the standard deviation in $T'$ and $Ca$ for all >200 measurements within one rearrangement zone at one $Q$ value, respectively (see the previous section and Fig. 4). The three dashed lines that follow the data points have logarithmic slopes of -0.14, -0.61, and -0.37. The two vertical dashed lines indicate the transition between between Regimes 1 and 2 at $Ca_{1-2}$, and between Regimes 2 and 3 at $Ca_{2-3}$. **(d)** $dT''$, $T'$, and $De$ as a function of $Ca$. $Ca_{S-L}$ represent the transition the solid-like to liquid-like transition when $De = 1$. Reproduced from [81].

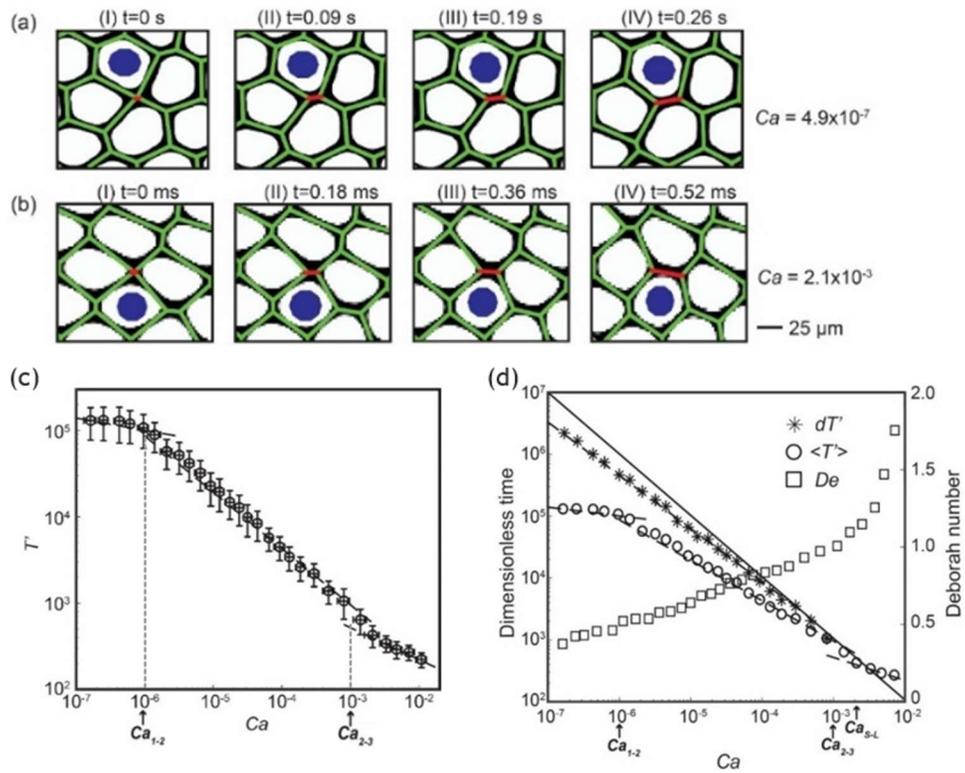

Figure 7. **(a)-(c)** Trajectories of 2000 droplets advected through the channel for: (a) $1.1 \times 10^{-7} < Ca < 1.6 \times 10^{-6}$ (within Regime 1), (b) $8.5 \times 10^{-6} < Ca < 6.3 \times 10^{-5}$ (within Regime 2), and (c) $7.9 \times 10^{-4} < Ca < 8.5 \times 10^{-3}$ (within Regime 3), respectively. Each marker represents the instantaneous position of the centroid of one droplet. The insets are time-averaged, x-component of droplet velocity ($\overline{U_x}$) across the width of the channel calculated at the locations marked by the red and green boxes, respectively. $\overline{U_x}$ is normalized to $\overline{U_x}_{max}$, the maximum x-component of velocity observed in the constriction. The rearrangement zones are indicated by blue arrows in (a) and (b). **(d)-(f)** Snapshots of successive T1 events at: (a) $\langle Ca \rangle = 2.5 \times 10^{-7}$ (Regime 1), (b) $\langle Ca \rangle = 1.2 \times 10^{-5}$ (Regime 2), and (c) $\langle Ca \rangle = 1.4 \times 10^{-3}$ (Regime 3), respectively. The blue arrows indictae the direction along which successive T1 events occur. For T1 events along the 60 degree slip plane, the converging (diverging) drops are colored in green (red). For T1 events along the direction of imposed flow, the converging (diverging) drops are colored in yellow (purple). For both (d) and (e), each sequence shows 5 successive T1 events along the slip plane. For (f), the sequence shows 5 successive T1 events along the slip plane and the flow direction, respectively. Reproduced from [81].

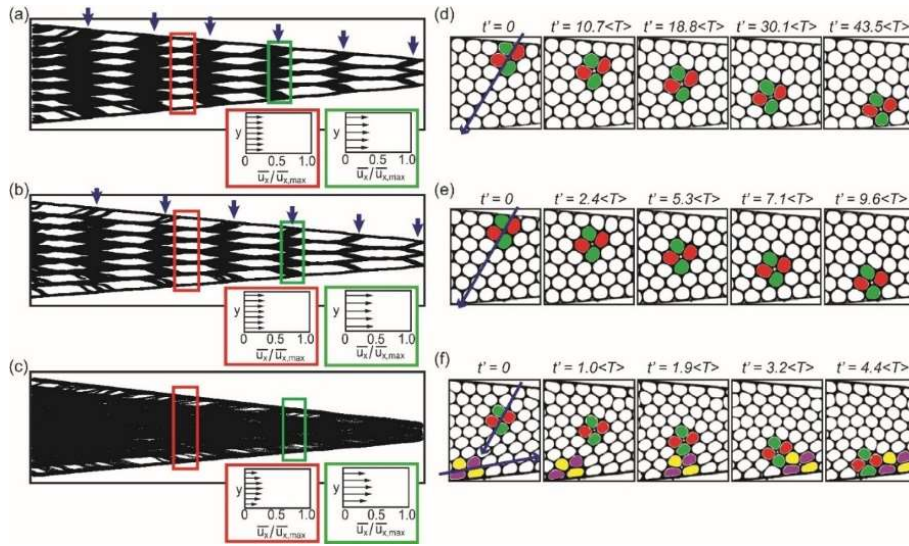

Figure 8. **(a)** Images of the flow of a concentrated emulsion consisting of 40-pL drops in a channel with a constriction. The average velocities of the flow at the constriction were (i) $v$ = 6.2 cm/s, and (ii) $v$ = 64.8 cm/s respectively. The blue and red boxes correspond to regions where we tracked drops upstream and downstream. **(b)** Histograms showing the frequency of occurrence of droplet sizes upstream (blue) and downstream (red) of the constriction, with plots (i) and (ii) corresponding to images (i) and (ii) in (a). **(c)** Breakup fraction as a function of capillary number at different drop sizes and constriction geometries. **(d)** Breakup fraction as a function of capillary number for emulsions with different viscosity ratios. A1-A8 in (d) represent experiments with different combinations of droplet size and channel constriction confinement (defined by constriction width and height), and B1-B5 in (d) represent experiments with various viscosity ratios. **(e)** Breakup fraction as a function of the product of capillary number, viscosity ratio, and confinement factor. The dashed lines are for visual guides only. A1-A8 in (a) represent experiments with different combinations of droplet size and channel constriction confinement (defined by constriction width and height), and B1-B5 in (b) represent experiments with various viscosity ratios. $Ca = \frac{\mu_c Gr}{\sigma}$, where $\mu_c$ is the dynamic viscosity of the continuous phase, $G$ is the strain rate in the constriction. See our previous work for detail [20,71]. Reproduced from [20,71] with permission from the Royal Society of Chemistry.

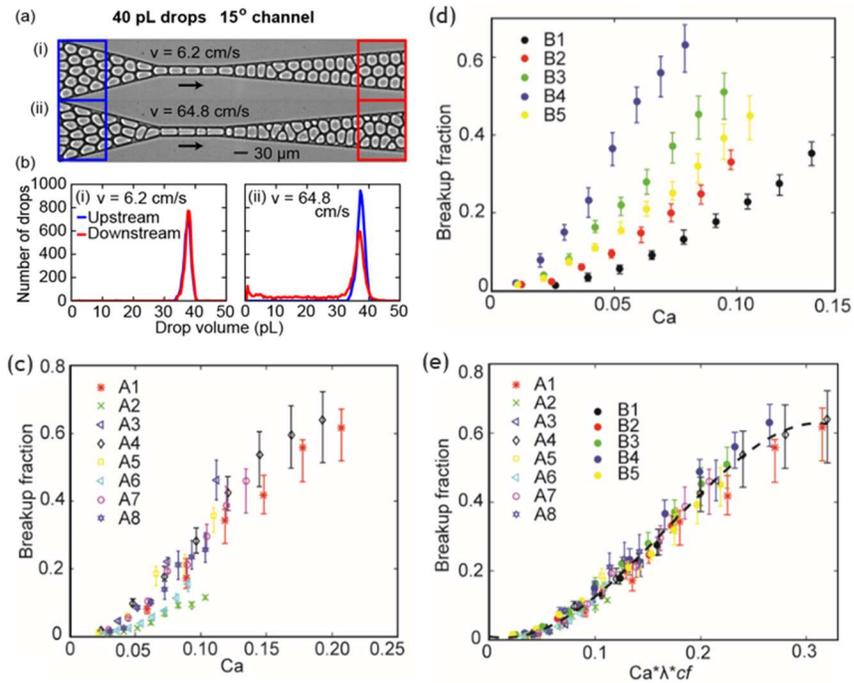

Figure 9. **(a)** Scheme of the microchannel used in our experiments with two possible outcomes of a pair of droplets entering the constriction. The initial offset Δx determines if breakup occurs. Only 3 droplets of the concentrated emulsion are shown. **(b)** Breakup regime map of normalized droplet pair offset as a function of $Ca$. **(c)** Snapshots of drop pairs in (i) Region I, (ii) Region II (break), (iii) Region II (no break), and (iv) Region III at $Ca = 0.014$. The leading-edge offset is (i) 3.54 μm, (ii) 28.32 μm, (iii) 28.32 μm, and (iv) 34.22 μm, respectively. Reproduced from [91], with the permission of AIP Publishing

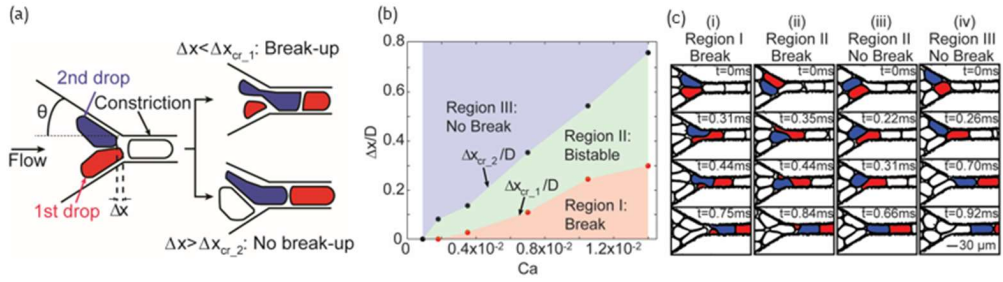

Figure 10. **(a)** NP droplet breakup fraction as a function of the product of capillary number ($Ca$) and confinement factor ($cf$). The black dashed line is a visual guide. The gray dashed line is the visual guide adapted from surfactant droplets (see Fig. 8e). **(b)** Effect of viscosity ratio on breakup of NP emulsion droplets. Breakup fraction as a function of capillary number at different viscosity ratios λ: A5 (λ=0.78), B1 (λ=1.38), B2 (λ=2.90), and B3 (λ=17.04). The dashed line is for visual guide only. Reproduced from [10], with the permission of AIP Publishing

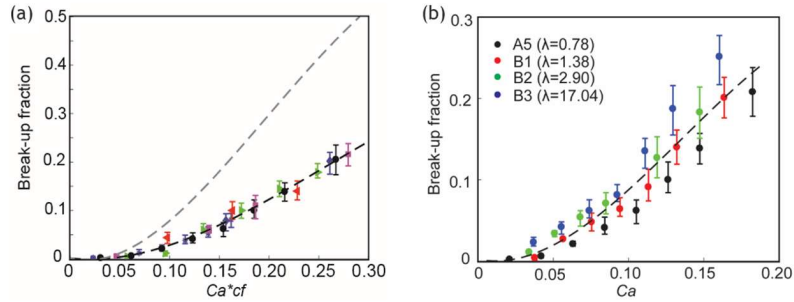

Figure 11. **(a)** Schematic diagram of the channel geometry. **(b)** Microscopic image of the emulsion flowing in the channel. The image was converted to a binary image to increase the contrast of droplet borders. **(c)** Breakup fraction as a function of volume fraction at different $Q_{total}$ or Ca, and **(d)** as a function of $Q_{total}$ or Ca at various volume fractions. The curves are fits from the data. In **(c)-(d)**, each data point and the corresponding error bar are obtained from N = 4,000 – 21,000 drops. **(e)** Droplet throughput as a function of volume fraction at different breakup fractions. See our work for a detailed description of data processing [96]. Reproduced from [96], with the permission of AIP Publishing

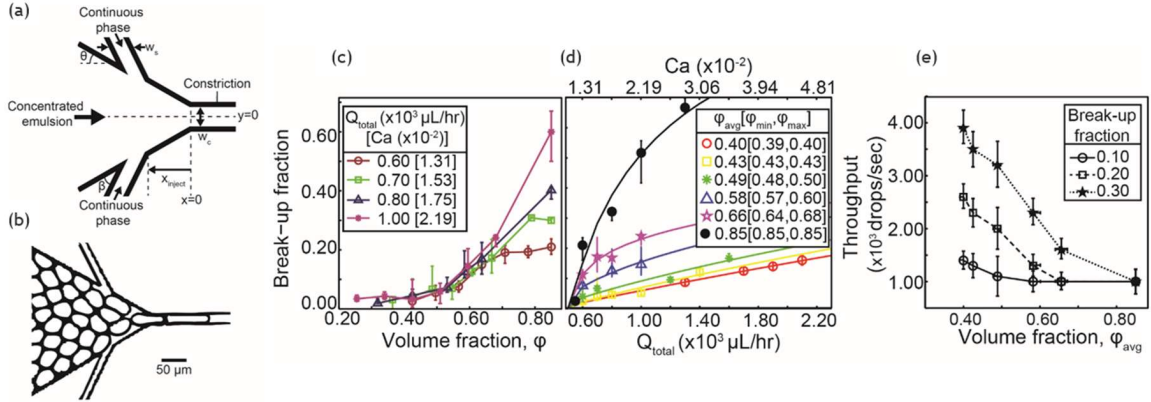

Figure 12. **(a)** Schematic diagram of the channel geometry. **(b)** Drop breakup fraction in the presence of an obstacle normalized to that in the absence of the obstacle ($\beta$) as a function of normalized obstacle-to-constriction distance, $x/D_d$ for various obstacle diameters ($D_o$) and capillary numbers Ca. Datasets F#.5 (where # = 1-6), G#.5 (where # = 1-10), I#.5 (where # = 1-10) have $D_o$ = 90 µm, 112.5 µm, and 150 µm, respectively but the same Ca = $10.8 \times 10^{-3} \pm 0.5 \times 10^{-3}$. Datasets G#.3 (where # = 4-7), G#.4 (where # = 1-9), and G#.5 (where # = 1-10) have the same $D_o$ = 112.5 µm but different Ca = $7.6 \times 10^{-3} \pm 0.6 \times 10^{-3}$, $9.3 \times 10^{-3} \pm 0.6 \times 10^{-3}$, and $10.8 \times 10^{-3} \pm 0.5 \times 10^{-3}$, respectively. See [107] for details of the experiments. **(c)** Regime map of $\beta$ at various $l_{pow}/D_d$ and $l_{ow}/D_d$ values. All data shown here were measured from channels with a half angle $\theta = 30°$ and Ca ranging from $5.46 \times 10^{-3}$ to $13.1 \times 10^{-3}$. Regions 0 to V are defined in the map. The red, black, and green markers indicate cases with $\beta > 1.0$, $0.1 < \beta \leq 1.0$, and $\beta \leq 0.1$, respectively. The green "+" marker indicates the case with minimum $\beta = 1.2 \times 10^{-3}$. The light gray area indicates geometries that were inaccessible experimentally as $l_{ow}$ cannot exceed $l_{pow}$. **(d)** Snapshots of the emulsion flowing in representative channel geometries in regions 0 to V. The circle filled with blue hashed lines is the obstacle. The scale bar is 100 µm. **(e)** Time series showing that drops alternate in an orderly manner and enter the constriction one at a time in a channel with an optimally-placed obstacle. Reproduced from [107].

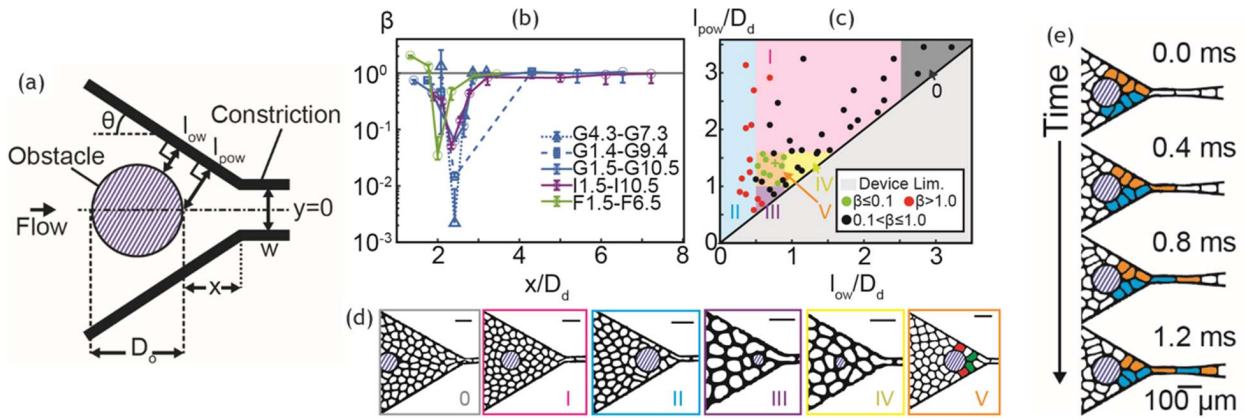

Figure 13. **(a)** Double emulsion formation through hierarchical flow-focusing microchannel obtained via VOF-finite volume approach [144]. **(b)** Numerical simulation of droplets assembly in a divergent channel. Simulation made with Aphros software [125]. **(c)** BIM simulation of droplets passing through a constriction [119]. **(d)** Droplets generation in a microfluidic T-junction via VOF simulations [123].

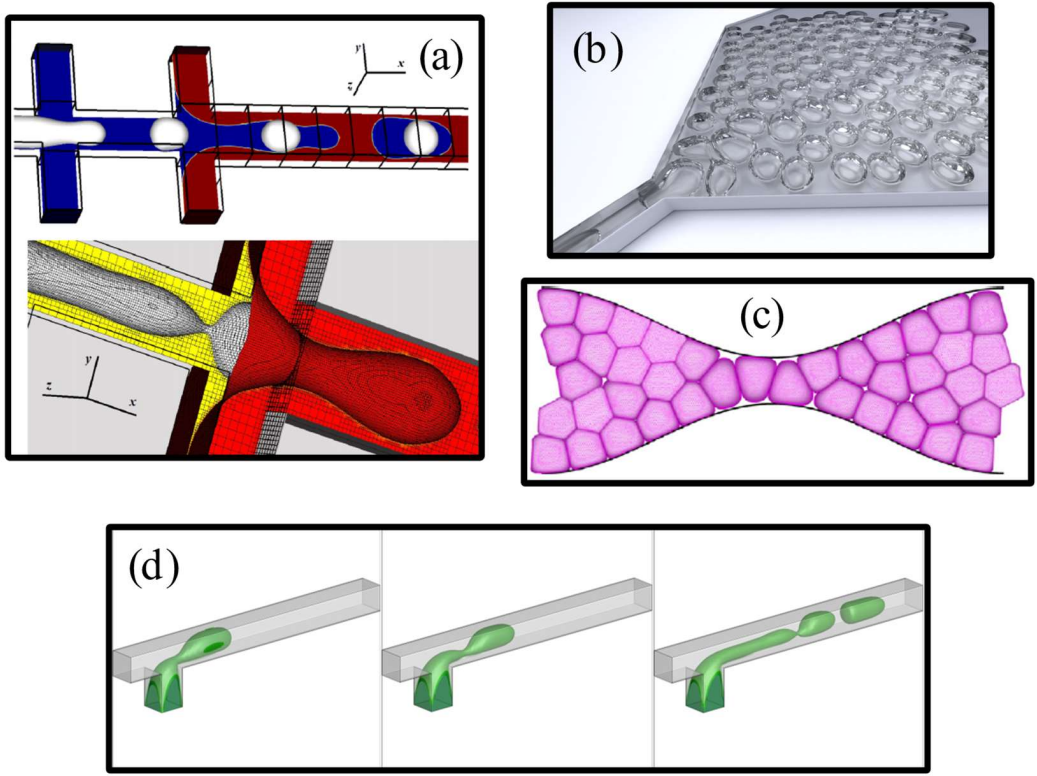

Figure 14. **(a)** A snapshot of the droplets in a concentrated emulsion, flowing from left to right. The upper panel corresponds to experiment, and lower to LB simulation [145]. **(b)** Soft flowing crystals formation within a microfluidic focuser: LB color-gradient simulations versus experiments [110]. **(c)** Deformation and breakup dynamics of droplets within a tapered channel: LB color-gradient simulations vs experiments [142]. **(d)** Jet-break up in soft granular materials flowing in a microfluidic focuser: LB color-gradient simulations vs experiments [147]. **(e)** Free-energy LB simulations of dense emulsions flowing in a straight channel [148].

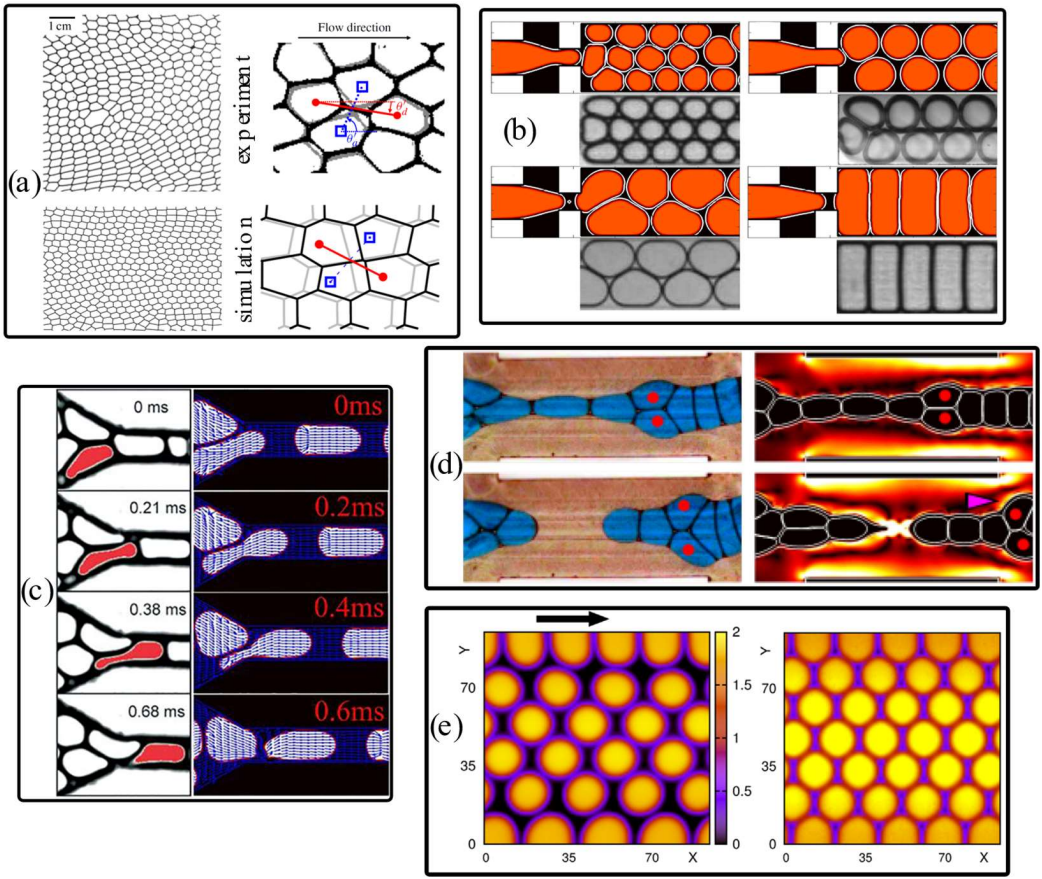